\documentclass[aps,prc,preprint,superscriptaddress,nofootinbib]{revtex4-1}
\usepackage{graphicx}% Include figure files
\usepackage{dcolumn}% Align table columns on decimal point
\usepackage{bm}% bold math
\usepackage{epstopdf}
\usepackage{amsmath}
\usepackage{hyperref}% add hypertext capabilities
\usepackage{color}
\usepackage{cancel}

\begin{document}

%\begin{CJK*}{UTF8}{}
\title{Generalized time-dependent generator coordinate method for induced fission dynamics}
%\CJKfamily{gbsn}
\author{B. Li}
\affiliation{State Key Laboratory of Nuclear Physics and Technology, School of Physics, Peking University, Beijing 100871, China}
\author{D. Vretenar}
\email{vretenar@phy.hr}
\affiliation{Physics Department, Faculty of Science, University of Zagreb, 10000 Zagreb, Croatia}
\affiliation{State Key Laboratory of Nuclear Physics and Technology, School of Physics, Peking University, Beijing 100871, China}
\author{T. Nik\v si\' c}
\affiliation{Physics Department, Faculty of Science, University of Zagreb, 10000 Zagreb, Croatia}
\affiliation{State Key Laboratory of Nuclear Physics and Technology, School of Physics, Peking University, Beijing 100871, China}
\author{J. Zhao}
\affiliation{Center for Circuits and Systems, Peng Cheng Laboratory, Shenzhen 518055, China}
\author{P. W. Zhao}
\email{pwzhao@pku.edu.cn}
\affiliation{State Key Laboratory of Nuclear Physics and Technology, School of Physics, Peking University, Beijing 100871, China}
\author{J. Meng}
\email{mengj@pku.edu.cn}
\affiliation{State Key Laboratory of Nuclear Physics and Technology, School of Physics, Peking University, Beijing 100871, China}

\begin{abstract}
The generalized time-dependent generator coordinate method (TD-GCM) is extended to include pairing correlations. The correlated GCM nuclear wave function is expressed in terms of time-dependent generator states and weight functions. The particle-hole channel of the effective interaction is determined by a Hamiltonian derived from an energy density functional, while pairing is treated dynamically in the standard BCS approximation with time-dependent pairing tensor and single-particle occupation probabilities. With the inclusion of pairing correlations, various time-dependent phenomena in open-shell nuclei can be described more realistically. The model is applied to the description of saddle-to-scission dynamics of induced fission. The generalized TD-GCM charge yields and total kinetic energy distribution for the fission of $^{240}$Pu, are compared to those obtained using the standard time-dependent density functional theory (TD-DFT) approach, and with available data.

{\bf keywords}: nuclear density functional theory, generator coordinate method, fission dynamics
\end{abstract}

\date{\today}

\maketitle

%\end{CJK*}
%--------------

%-------------------------------------------------------------------------------------------------------------
\section{Introduction}
%-------------------------------------------------------------------------------------------------------------

The process of spontaneous or induced fission, by which an atomic nucleus breaks into two or more fragments, with the  corresponding release of an enormous amount of energy, has become one of the best studied phenomena in nuclear physics. From a modern perspective, the fission process is particularly important as a representative case of large-amplitude collective motion in a self-bound mesoscopic system, that exhibits both classical and quantal characteristics. Fission is also relevant for the stability of superheavy elements, production of short-lived exotic nuclides far from stability, nuclear astrophysics and the mechanism of nucleosynthesis.
A wealth of experimental results on nuclear fission have been accumulated over more than eight decades, and a basic understanding of the mechanism gained. Several very successful phenomenological methods have been developed that reproduce, to various degrees of accuracy, low- and medium-energy fission observables. A unified microscopic framework for the description of the entire fission process, from the quasi-stationary initial state to the outer fission barrier (saddle point), and the complicated dissipative dynamics that lead to the emergence of excited fragments at scission, remains a formidable challenge for nuclear theory.

In the last decade significant advances have been made in the development of microscopic models of induced fission dynamics. The two basic approaches are based on the time-dependent generator coordinate method (TD-GCM) and the
time-dependent density functional theory (TD-DFT), respectively. In the former, the nuclear wave function is represented by a superposition of generator states that are functions of collective coordinates, and can be applied to an adiabatic description of the entire fission process \cite{krappe12,schunck16,younes19,Regnier2016_PRC93-054611,Verriere2020_FP8-233,Tao2017PRC,Zhao2019PRC_2}. This approach is fully quantum mechanical but only takes into account collective degrees of freedom and, therefore, cannot be used to describe the highly dissipative dynamics that occurs beyond the saddle point.
While the TD-DFT framework consistently incorporates the one-body dissipation mechanism, it can only model a single fission event by propagating nucleons independently toward scission and beyond \cite{simenel18,nakatsukasa16,stevenson19,bulgac16,magierski17,scamps18,bulgac19,bulgac20,Ren2022}. In the fission case, TD-DFT models the classical evolution of independent nucleons in mean-field potentials, and cannot be applied in the classically forbidden region of the collective space nor does it take into account quantum fluctuations.

In Ref.~\cite{Li2023PRC}, an implementation of the generalized TD-GCM has been developed and applied to the dynamics of small and large amplitude collective motion of atomic nuclei. Both the generator states and weight functions of the GCM correlated wave function depend on time. The initial generator states are obtained as solution of deformation-constrained self-consistent mean-field equations, and are evolved in time by the standard mean-field equations of nuclear density functional theory (TD-DFT). The TD-DFT trajectories are used as a basis in which the TD-GCM wave function is expanded. As a first application, in Ref.~\cite{Li2023PRC} the generalized TD-GCM has been employed in a study of excitation energies and spreading width of giant resonances, and in an initial exploration of the dynamics of induced fission. An equivalent realization of the model has very recently been applied to a study of collective multiphonon states in nuclei \cite{Marevic_2023}.

In this work, the generalized TD-GCM is extended with the inclusion of dynamical pairing correlations and employed in a systematic investigation of induced fission dynamics. Pairing is included in the TD-DFT framework in a standard BCS approximation, with time-dependent pairing tensor and occupation probabilities. A basis of TD-DFT fission trajectories, generally non-orthogonal and overcomplete, is used to build the correlated TD-GCM wave function. With the coherent superposition of TD-DFT trajectories in the generalized TD-GCM, fission dynamics is described fully quantum mechanically in an approach that extends beyond the adiabatic approximation of the standard GCM and, simultaneously, includes quantum fluctuations.

In Sec.~\ref{sec_theo}, we outline the theoretical framework of generalized TD-GCM that also takes into account pairing correlations. Section~\ref{sec_numerics} collects the numerical details of the fission calculation. As an application of the generalized TD-GCM, the dynamics of induced fission of $^{240}$Pu is discussed in Sec.~\ref{sec_240Pu}. Section~\ref{sec_summ} summarizes the results and presents a brief outlook for future studies.
%-----------------------------------------------------------------------------------------------------------------------------
\section{Theoretical framework: Generalized time-dependent GCM with pairing interactions}\label{sec_theo}
%-----------------------------------------------------------------------------------------------------------------------------

The Griffin-Hill-Wheeler (GHW) ansatz for the TD-GCM correlated nuclear wave function reads~\cite{Reinhard1983NPA,Regnier2019PRC,Verriere2020_FP8-233}
\begin{equation}
    |\Psi(t)\rangle=\int_{\bm q} d{\bm q}~f_{\bm q}(t) |\Phi_{\bm q}(t)\rangle,
    \label{GHW_wf}
\end{equation}
where the vector ${\bm q}$ denotes the continuous real {\em generator coordinates} that parametrize the collective degrees of freedom. This wave function is a linear superposition of, generally non-orthogonal, many-body {\em generator states} $|\Phi_{\bm q}(t)\rangle$, and $f_{\bm q}(t)$ are the corresponding complex-valued {\em weight functions}.
The generalized TD-GCM without the inclusion of pairing correlations has been implemented in the first part of this work \cite{Li2023PRC}. In this study, pairing is also taken into account, and the discretized generator coordinates are the mass multipole moments (axial quadrupole and octupole) of the nucleon density distribution. Thus, the nuclear wave function
\begin{equation}
    |\Psi(t)\rangle=\sum_{\bm q} f_{\bm q}(t) |\Phi_{\bm q}(t)\rangle,
    \label{Eq_collec_wfs}
\end{equation}
is the solution of the time-dependent equation
\begin{equation}
    i\hbar\partial_t|\Psi(t)\rangle=\hat{H}|\Psi(t)\rangle,
    \label{Eq_td_eq}
\end{equation}
where $\hat{H}$ is the Hamiltonian of the nuclear system. From a time-dependent variational principle~\cite{Regnier2019PRC}, one obtains the equation of motion for the weight functions
\begin{equation}
     i\hbar \mathcal{N}\dot{f}=(\mathcal{H}-\mathcal{H}^{MF})f,
    \label{TD-HW-f}
\end{equation}
which, in the discretized collective space, reads
\begin{equation}
\sum_{\bm q} i\hbar \mathcal{N}_{\bm{q'q}}(t)\partial_t f_{\bm q}(t)+\sum_q \mathcal{H}_{\bm {q'q}}^{MF}(t) f_{\bm q}(t) =\sum_q \mathcal{H}_{\bm {q'q}}(t)f_{\bm q}(t).
\end{equation}
The time-dependent kernels
\begin{subequations}
 \begin{align}
&\mathcal{N}_{\bm{q'q}}(t)=\langle\Phi_{\bm {q'}}(t)|\Phi_{\bm q}(t)\rangle,\label{Eq_N}\\
&\mathcal{H}_{\bm{q'q}}(t)=\langle\Phi_{\bm {q'}}(t)|\hat{H}|\Phi_{\bm q}(t)\rangle,\label{Eq_H}\\
&\mathcal{H}^{MF}_{\bm{q'q}}(t)=\langle\Phi_{\bm {q'}}(t)|i\hbar\partial_t|\Phi_{\bm q}(t)\rangle,
\label{Eq_H_mf}
 \end{align}
\end{subequations}
include the overlap, the Hamiltonian, and the time derivative of the generator states, respectively.
%-----------------------------------------------------------------------------------------------
\subsection{Time-dependent quasiparticle vacuum  $|\Phi_{\bf q}(t)\rangle$}
%-----------------------------------------------------------------------------------------------

The evolution in time of the quasiparticle vacuum characterized by a vector of generator coordinates ${\bm q}$
\begin{equation}
   |\Phi_{\bm q}(t)\rangle = \prod_{k>0}[\mu_{{\bm q},k}(t)+\nu_{{\bm q},k}(t)c_{{\bm q},k}^\dagger(t)c_{{\bm q},\bar{k}}^\dagger(t)]|-\rangle \;,
\label{Eq_Slater}
\end{equation}
is modeled by the time-dependent covariant density functional theory~\cite{Ren2022,Ren2022a}, using the time-dependent BCS approximation~\cite{ebata10TDBCS,Scamps13TDBCS}.
In Eq.(\ref{Eq_Slater}), $\mu_{{\bm q},k}(t)$ and $\nu_{{\bm q},k}(t)$ are the parameters of the transformation between the canonical and quasiparticle bases, and $c_{{\bm q},k}^\dagger(t)$ denotes the creation operator associated with the canonical state $\phi_k^{\bm q}(\bm{r},t)$.~The time evolution of $\phi_k^{\bm q}(\bm{r},t)$ is determined by the time-dependent Dirac equation
\begin{equation}\label{Eq_td_Dirac_eq_BCS}
  i\frac{\partial}{\partial t}\phi_k^{\bm q}(\bm{r},t)=[\hat{h}^{\bm q}(\bm{r},t)-\varepsilon_k^{\bm q}(t)]\phi_k^{\bm q}(\bm{r},t),
\end{equation}
where $\varepsilon_k^{\bm q}(t)=\langle\psi_k^{\bm q}|\hat{h}^{\bm q}|\psi_k^{\bm q}\rangle$ is the single-particle energy, and the single-particle Hamiltonian $\hat{h}^{\bm q}(\bm{r},t)$ reads
\begin{equation}
   \hat{h}^{\bm q}(\bm{r},t) = \bm{\alpha}\cdot(\hat{\bm{p}}-\bm{V}_{\bm q})+V^0_{\bm q}+\beta(m_N+S_{\bm q}).
   \label{Ham_D}
\end{equation}
Here, $m_N$~is the nucleon mass, and the scalar $S_{\bm q}(\bm{r},t)$ and four-vector $V^{\mu}_{\bm q}(\bm{r},t)$ potentials at every instant are determined by the time-dependent densities and currents in the isoscalar-scalar, isoscalar-vector, and isovector-vector channels.
In the present study, we employ the point-coupling relativistic energy density functional PC-PK1 \cite{Zhao2010PRC}, and the explicit expressions for the potentials read
\begin{subequations}
  \begin{align}
    S_{\bm q}(\bm{r})=\,&\alpha_S\rho_S^{\bm q}+\beta_S(\rho_S^{\bm q})^2+\gamma_S(\rho_S^{\bm q})^3+\delta_S\Delta\rho_S^{\bm q},\\
    V^\mu_{\bm q}(\bm{r})=\,&\alpha_Vj^{{\bm q},\mu}+\gamma_V(j^{{\bm q},\mu} j^{\bm q}_\mu)j^{{\bm q},\mu}+\delta_V\Delta j^{{\bm q},\mu}+\tau_3\alpha_{TV}j_{TV}^{{\bm q},\mu}+\tau_3\delta_{TV}\Delta j_{TV}^{{\bm q},\mu}+e\frac{1-\tau_3}{2}A^{{\bm q},\mu},
  \end{align}
\end{subequations}
where $\tau_3$ is the isospin Pauli matrix, and $A^{{\bm q},\mu}$ is the electromagnetic vector potential.
The densities and currents are defined in terms of occupied single-particle wave functions~$\phi_k^{\bm q}(\bm{r},t)$ and the occupation probabilities $n_{{\bm q},k}(t)$:
\begin{subequations}\label{Eq_density_current}
  \begin{align}
    &\rho_S^{\bm q}(\bm{r},t)=\sum_k^{l_q} n_{{\bm q},k}(t)\bar{\phi}^{\bm q}_k(\bm{r},t)\phi^{\bm q}_k(\bm{r},t),\\
    &j^{{\bm q},\mu}(\bm{r},t)=\sum_k^{l_q} n_{{\bm q},k}(t)\bar{\phi}^{\bm q}_k(\bm{r},t)\gamma^\mu\phi^{\bm q}_k(\bm{r},t),\\
    &j_{TV}^{{\bm q},\mu}(\bm{r},t)=\sum_k^{l_q}  n_{{\bm q},k}(t)\bar{\phi}^{\bm q}_k(\bm{r},t)\gamma^\mu\tau_3\phi^{\bm q}_k(\bm{r},t),
  \end{align}
\end{subequations}
where $l_q$ is the number of the canonical basis states.
The time evolution of the occupation probability $n_{{\bm q},k}(t)=|\nu_{{\bm q},k}(t)|^2$, and pairing tensor $\kappa_{{\bm q},k}(t)=\mu_{{\bm q},k}^{*}(t)\nu_{{\bm q},k}(t)$, is governed by the following equations:
\begin{subequations}\label{Eq_td_nkapp_eq_BCS}
   \begin{align}
     &i\frac{d}{dt}n_{{\bm q},k}(t)=\kappa_{{\bm q},k}(t)\Delta_{{\bm q},k}^*(t)-\kappa_{{\bm q},k}^{*}(t)\Delta_{{\bm q},k}(t),\\
     &i\frac{d}{dt}\kappa_{{\bm q},k}(t)=[\varepsilon_k^{\bm q}(t)+\varepsilon_{\bar{k}}^{\bm q}(t)]\kappa_{{\bm q},k}(t)+\Delta_{{\bm q},k}(t)[2n_{{\bm q},k}(t)-1] ,
   \end{align}
\end{subequations}
(for details, see Refs.~\cite{Scamps13TDBCS,ebata10TDBCS}). In time-dependent calculations, a monopole pairing interaction is employed, and the gap parameter $\Delta_{{\bm q},k}(t)$ is defined in terms of single-particle energies and the pairing tensor,
\begin{equation}
  \Delta_{{\bm q},k}(t)=\left[G\sum_{k'>0}f(\varepsilon_{k'}^{\bm q})\kappa_{{\bm q},k'}\right]f(\varepsilon_k^{\bm q}),
\end{equation}
where $f(\varepsilon_k^{\bm q})$ is the cut-off function for the pairing window \cite{Scamps13TDBCS}, and $G$ is the pairing strength.

%-------------------------------------------------------------------------------
\subsection{Overlap kernel $\mathcal{N}_{\bf {q'q}}(t)$}
%-------------------------------------------------------------------------------

According to Eq.(\ref{Eq_Slater}), the expression for the overlap kernel Eq.(\ref{Eq_N}) can be written in the following form:
\begin{equation}\label{Eq_norm_kernel}
    \begin{split}
    \mathcal{N}_{\bm {q'q}}(t)&=\langle\Phi_{\bm {q'}}(t)|\Phi_{\bm q}(t)\rangle\\
    &=\langle-|\prod_{k'>0}[\mu_{{\bm q'},k'}^*(t)+\nu_{{\bm q'},k'}^*(t)c_{{\bm q'},\bar{k}'}(t)c_{{\bm q'},k'}(t)]
    \prod_{k>0}[\mu_{{\bm q},k}(t)+\nu_{{\bm q},k}(t)c_{{\bm q},k}^\dagger(t)c_{{\bm q},\bar{k}}^\dagger(t)]|-\rangle\\
    &=\frac{(-1)^{(l_{\bm q'}-1)l_{\bm q'}/2}}{\prod_{k'}^{l_{\bm q'}/2} \prod_k^{l_{\bm q}/2} \nu_{{\bm q'},k'}^{*}\nu_{{\bm q},k}}\langle-|\beta_{{\bm q'},1}^\dagger...\beta_{{\bm q'},l_{\bm q'}}^\dagger\beta_{{\bm q},1}...\beta_{{\bm q},l_{\bm q}}|-\rangle, \\
    \end{split}
\end{equation}
where $\beta^\dagger_{{\bm q},k}$ is the quasi-particle creation operator associated with the quasiparticle vacuum $|\Phi_{\bf q}(t)\rangle$.
The overlap between two quasi-particle vacuums can be calculated using the Pfaffian algorithms
developed in Refs.~\cite{Robledo2009PRC,Hu2014PLB}.

%-------------------------------------------------------------------------------
\subsection{Energy kernel $\mathcal{H}_{\bf {q'q}}(t)$}
%-------------------------------------------------------------------------------

 For the point-coupling relativistic energy density functional PC-PK1 \cite{Zhao2010PRC},
one obtains the expression for the energy kernel $\mathcal{H}^{\rm DF}(t)$, under the assumption \cite{nakatsukasa16} that it only depends on the transition densities at time $t$:
\begin{equation}
\begin{aligned}
        \mathcal{H}^{{\rm DF}}_{\bm {q'q}}(t)=\langle\Phi_{\bm {q'}}(t)|\hat{H}^{{\rm DF}}|\Phi_{\bm q}(t)\rangle&=\langle\Phi_{\bm {q'}}(t)|\Phi_{\bm q}(t)\rangle\cdot\int d^3r~\{ \rho_{\rm kin}(\bm{r},t)\\
        &+\frac{\alpha_S}{2}\rho_S(\bm{r},t)^2+\frac{\beta_S}{3}\rho_S(\bm{r},t)^3\\
        &+\frac{\gamma_S}{4}\rho_S(\bm{r},t)^4+\frac{\delta_S}{2}\rho_S(\bm{r},t)\Delta\rho_S(\bm{r},t)\\
        &+\frac{\alpha_V}{2} j^\mu(\bm{r},t) j_\mu(\bm{r},t)+\frac{\gamma_V}{4}(j^\mu(\bm{r},t) j_\mu(\bm{r},t))^2\\
        &+\frac{\delta_V}{2} j^\mu(\bm{r},t)\Delta j_\mu(\bm{r},t)+\frac{\alpha_{TV}}{2} j^\mu_{TV}(\bm{r},t)\cdot [j_{TV}(\bm{r},t)]_\mu\\
        &+\frac{ \delta_{TV}}{2} j^\mu_{TV}(\bm{r},t)\cdot \Delta[j_{TV}(\bm{r},t)]_\mu+\frac{e^2}{2}j^\mu_p(\bm{r},t) A_\mu(\bm{r},t) \},
\end{aligned}
\end{equation}
where the densities and currents~$\rho_{\rm kin}$,~$\rho_S$,~$j^\mu$,~$j_{TV}^\mu$,~and~$j_p^\mu$ read
\begin{subequations}
 \begin{align}
        &\rho_{\rm kin}(\bm{r},t)=\sum_{k'}^{l_{\bm q'}}\sum_{k}^{l_{\bm q}}\bar{\phi}_{k'}^{\bm {q'}}(\bm{r},t)(-i\bm{\gamma}\cdot\bm{\nabla}+m_N)\phi_{k}^{\bm q}(\bm{r},t)\rho^{\rm tran}_{{\bm {q'q}},k'k}(t),\\
        &\rho_S(\bm{r},t)=\sum_{k'}^{l_{\bm q'}}\sum_{k}^{l_{\bm q}}\bar{\phi}_{k'}^{\bm {q'}}(\bm{r},t)\phi_{k}^{\bm q}(\bm{r},t)\rho^{\rm tran}_{{\bm {q'q}},k'k}(t),\\
        &j^\mu(\bm{r},t)=\sum_{k'}^{l_{\bm q'}}\sum_{k}^{l_{\bm q}}\bar{\phi}_{k'}^{\bm {q'}}(\bm{r},t)\gamma^\mu\phi_{k}^{\bm q} (\bm{r},t)\rho^{\rm tran}_{{\bm {q'q}},k'k}(t),\\
        &j_{TV}^\mu(\bm{r},t)=\sum_{k'}^{l_{\bm q'}}\sum_{k}^{l_{\bm q}}\bar{\phi}_{k'}^{{\bm q'}}(\bm{r},t)\tau_3\gamma^\mu\phi_{k}^{\bm q}(\bm{r},t)\rho^{\rm tran}_{{\bm {q'q}},k'k}(t),\\
        &j_p^\mu(\bm{r},t)=\frac{1-\tau_3}{2}\sum_{k'}^{l_{\bm q'}}\sum_{k}^{l_{\bm q}}\bar{\phi}_{k'}^{\bm {q'}}(\bm{r},t)\gamma^\mu\phi_{k}^{\bm q}(\bm{r},t)\rho^{\rm tran}_{{\bm {q'q}},k'k}(t).
\end{align}
\end{subequations}
The transition density matrix $\rho^{\rm tran}(t)$ is defined by the following relation
\begin{equation}
\rho^{\rm tran}_{{\bm {q'q}},k'k}(t)=\frac{ \langle\Phi_{\bm {q'}}(t)|c_{{\bm {q'}},k'}^{\dagger}(t)c_{{\bm q},k}(t)|\Phi_{\bm q}(t)\rangle}{\langle\Phi_{\bm {q'}}(t)|\Phi_{\bm q}(t)\rangle}
=\nu_{{\bm q'},k'}^{*}(t)\nu_{{\bm q},k}(t)\frac{ \langle\Phi_{\bm {q'}}(t)|\beta_{{\bm {q'}},\bar{k}'}(t)\beta_{{\bm q},\bar{k}}^{\dagger}(t)|\Phi_{\bm q}(t)\rangle}{\langle\Phi_{\bm {q'}}(t)|\Phi_{\bm q}(t)\rangle}.
\end{equation}
The numerator of the transition density matrix $\rho^{\rm tran}_{{\bm {q'q}},k'k}(t)$ is the overlap between a quasi-particle vacuum with $\left(l_{\bm q'}-1\right)$ quasi-particle levels and a quasi-particle vacuum with $\left(l_{\bm q}-1\right)$ quasi-particle levels.
It can be calculated using the  Pfaffian algorithms~\cite{Hu2014PLB,Robledo2009PRC}.

For the monopole pairing interaction, the pairing Hamiltonian operator $\hat{H}^{{\rm pair}}$ in 3D-lattice space is defined as
\begin{equation}
\hat{H}_{pair}=-\sum_{r_1,s_1>0,r_2,s_2>0} G(c^\dag_{r_1,s_1}c^\dag_{r_1,\bar{s}_1})(c_{r_2,\bar{s}_2}c_{r_2,{s}_2})
\end{equation}
where $c^\dag_{r_1,s_1}$ is the creation operator for the lattice coordinate wave function $|r_1,s_1\rangle$, and $r_1$ is the index of the lattice point and $s_1$ is the index of the spin.
One obtains the expression for the pairing part of energy kernel $\mathcal{H}^{\rm pair}(t)$ in the 3D-lattice space
\begin{equation}
\begin{aligned}
        &\mathcal{H}^{{\rm pair}}_{\bm {q'q}}(t)=\langle\Phi_{\bm {q'}}(t)|\hat{H}^{{\rm pair}}|\Phi_{\bm q}(t)\rangle\\
        &=- G ~\langle\Phi_{\bm {q'}}(t)|\Phi_{\bm q}(t)\rangle \sum_{k_1,k_2,k_3,k_4>0}[f(\varepsilon_{k_1}^{\bm q'})f(\varepsilon_{k_2}^{\bm q})f(\varepsilon_{k_3}^{\bm q'})f(\varepsilon_{k_4}^{\bm q})]^{1/2}  \\
        &\times \langle\phi^{\bm q'}_{k_1}|\phi^{\bm q}_{\bar{k}_2}\rangle\langle\phi^{\bm q'}_{\bar{k}_3}|\phi^{\bm q}_{k_4}\rangle
         [\kappa_{\bm{q'q},k_2k_1}^{\rm tran}(t)]^{*}\kappa_{\bm{qq'},k_3k_4}^{\rm tran}(t),\\
\end{aligned}
\end{equation}
where the transition pairing tensor matrix $\kappa^{\rm tran}(t)$ is defined by the following relation
\begin{equation}
\kappa^{\rm tran}_{{\bm {q'q}},k'k}(t)=\frac{ \langle\Phi_{\bm {q'}}(t)|c_{{\bm {q'}},k'}(t)c_{{\bm q},k}(t)|\Phi_{\bm q}(t)\rangle}{\langle\Phi_{\bm {q'}}(t)|\Phi_{\bm q}(t)\rangle}
=\mu_{{\bm q'},k'}^{*}(t)\nu_{{\bm q},k}(t)\frac{ \langle\Phi_{\bm {q'}}(t)|\beta_{{\bm {q'}},k'}(t)\beta_{{\bm q},\bar{k}}^{\dagger}(t)|\Phi_{\bm q}(t)\rangle}{\langle\Phi_{\bm {q'}}(t)|\Phi_{\bm q}(t)\rangle},
\end{equation}

In the BCS model, $|\Phi_{\bm q}(t)\rangle$ is not an eigenstate of the neutron (proton) number operator $\hat{N}$ ($\hat{Z}$), and its expectation value in the collective wave function generally deviates from the desired neutron number $N_0$ (proton number $Z_0$). The method developed in Ref.~\cite{Bonche1990NPA} is used to correct for variations of the nucleon number. The energy kernel finally reads
\begin{equation}
\mathcal{H}_{\bm {q'q}}(t)= \mathcal{H}^{{\rm DF}}_{\bm {q'q}}(t)+ \mathcal{H}^{{\rm pair}}_{\bm {q'q}}(t)
-\lambda_{N}^{\bm {q'q}}(t)[\langle\Phi_{\bm {q'}}(t)|\hat{N}|\Phi_{\bm q}(t)\rangle-N_0]-\lambda_{Z}^{\bm {q'q}}(t)[\langle\Phi_{\bm {q'}}(t)|\hat{Z}|\Phi_{\bm q}(t)\rangle-Z_0],
\end{equation}
where $\lambda_{i}^{\bm {q'q}}(t)$ is defined as the average of the chemical potentials $\lambda_{i}^{\bm {q'}}(t)$ and $\lambda_{i}^{\bm {q}}(t)$,
\begin{equation}
\lambda_{i}^{\bm {q'q}}(t)=\frac{\lambda_{i}^{\bm {q'}}(t)+\lambda_{i}^{\bm {q}}(t)}{2},~~~i=N,Z.
\end{equation}

%-------------------------------------------------------------------------------
\subsection{Mean-field kernel $\mathcal{H}^{MF}_{\bf {q'q}}(t)$}
%-------------------------------------------------------------------------------
From the expression for the time evolution of $|\Phi_{\bm q}(t)\rangle$~\cite{Ren2022,Ren2022a},
\begin{equation}\label{MF_kernel}
 \begin{aligned}
    &i\hbar\partial_t|\Phi_{\bm q}(t)\rangle\\
    &=i\hbar\sum_{k>0}\{\partial_t[c_{{\bm q},k}^{\dagger}(t)c_{{\bm q},\bar{k}}^\dagger(t)]+\dot{\mu}_{{\bm q},k}(t)+\dot{\nu}_{{\bm q},k}(t)c_{{\bm q},k}^\dagger(t)c_{{\bm q},\bar{k}}^\dagger(t)\}
    \prod_{j\neq k,j>0}[\mu_{{\bm q},j}(t)+\nu_{{\bm q},j}(t)c_{{\bm q},j}^\dagger(t)c_{{\bm q},\bar{j}}^\dagger(t)]|-\rangle\\
    &=\sum_{k}^{l_q}[\hat{h}^{\bm q}(\bm{r},t)-\varepsilon_k^{\bm q}(t)]c_{{\bm q},k}^{\dagger}(t)c_{{\bm q},k}(t)|\Phi_{\bm q}(t)\rangle
   +i\hbar\sum_{k>0}\sqrt{|\dot{\mu}_{{\bm q},k}(t)|^2+|\dot{\nu}_{{\bm q},k}(t)|^2}~|\tilde{\Phi}_{{\bm q},k}(t)\rangle,
 \end{aligned}
\end{equation}
where the Slater determinant $|\tilde{\Phi}_{{\bm q},k}(t)\rangle$ is defined as
\begin{equation}
 \begin{aligned}
|\tilde{\Phi}_{{\bm q},k}(t)\rangle&=[ \frac{\dot{\mu}_{{\bm q},k}(t)}{\sqrt{|\dot{\mu}_{{\bm q},k}(t)|^2+|\dot{\nu}_{{\bm q},k}(t)|^2}}
                        +\frac{\dot{\nu}_{{\bm q},k}(t)}{\sqrt{|\dot{\mu}_{{\bm q},k}(t)|^2+|\dot{\nu}_{{\bm q},k}(t)|^2}}c_{{\bm q},k}^\dagger(t)c_{{\bm q},\bar{k}}^\dagger(t)]\\
                        &\cdot\prod_{j\neq k,j>0}[\mu_{{\bm q},j}(t)+\nu_{{\bm q},j}(t)c_{{\bm q},j}^\dagger(t)c_{{\bm q},\bar{j}}^\dagger(t)]|-\rangle
\end{aligned}
\end{equation}
Eq.(\ref{Eq_H_mf}) can be written in the form
\begin{equation}
 \begin{aligned}
\mathcal{H}^{MF}_{\bm {q'q}}(t)&=\langle\Phi_{\bm q'}(t)|i\hbar\partial_t|\Phi_{\bm q}(t)\rangle\\
&=\langle\Phi_{\bm q'}(t)|\sum_{k}^{l_q}[\hat{h}^{\bm q}(\bm{r},t)-\varepsilon_k^{\bm q}(t)]c_{{\bm q},{k}}^{\dagger}(t)c_{{\bm q},{k}}(t)|\Phi_{\bm q}(t)\rangle\\
&+i\hbar\sum_{k>0}\sqrt{|\dot{\mu}_{{\bm q},k}(t)|^2+|\dot{\nu}_{{\bm q},k}(t)|^2}~\langle\Phi_{\bm q'}(t)|\tilde{\Phi}_{{\bm q},k}(t)\rangle.
\end{aligned}
\end{equation}
By expanding $[\hat{h}^{\bm q}(\bm{r},t)-\varepsilon_k^{\bm q}(t)]c^{\dagger}_{{\bm q},k}(t)$ in a complete basis $ c^{\dagger}_{{\bm q'},k'}(t)$,
\begin{equation}
     [\hat{h}^{\bm q}(\bm{r},t)-\varepsilon_k^{\bm q}(t)]c^{\dagger}_{{\bm q},k}(t)=\sum_{k'}\langle \phi^{{\bm q'}}_{k'}(\bm{r},t)|[\hat{h}^{\bm q}(\bm{r},t)-\varepsilon_k^{\bm q}(t)]|\phi^{\bm q}_{k}(\bm{r},t)\rangle c^{\dagger}_{{\bm q'},k'}(t),
\end{equation}
one obtains for $\mathcal{H}_{\bm {q'q}}^{MF}(t)$ the expression
\begin{equation}
 \begin{aligned}
\mathcal{H}_{\bm{q'q}}^{MF}(t)&=\langle\Phi_{\bm q'}(t)|\Phi_{\bm q}(t)\rangle\cdot\sum_{k'}^{l_{\bm q'}}\sum_{k}^{l_{\bm q}}\langle \phi^{\bm q'}_{k'}(\bm{r},t)|[\hat{h}^{\bm q}(\bm{r},t)-\varepsilon_k^{\bm q}(t)]|\phi^{\bm q}_{k}(\bm{r},t)\rangle\rho^{\rm tran}_{k'k}(t)\\
                              &+i\hbar\sum_{k>0}\sqrt{|\dot{\mu}_{{\bm q},k}(t)|^2+|\dot{\nu}_{{\bm q},k}(t)|^2}~\langle\Phi_{\bm q'}(t)|\tilde{\Phi}_{{\bm q},k}(t)\rangle,
\end{aligned}
\end{equation}
where $\dot{\mu}_{{\bm q},k}(t)$ and $\dot{\nu}_{{\bm q},k}(t)$ can be derived from Eq.~(\ref{Eq_td_nkapp_eq_BCS}),
and $\langle\Phi_{\bm q'}(t)|\tilde{\Phi}_{{\bm q},k}(t)\rangle$  can be obtained by the Pfaffian algorithms~\cite{Hu2014PLB,Robledo2009PRC}.
%-------------------------------------------------------------------------------
\subsection{Collective wave function $g(t)$}
%-------------------------------------------------------------------------------
Equation (\ref{TD-HW-f}) is not a collective Schr\"odinger equation, and the weight function $f_{\bm q}(t)$ is not a probability amplitude of finding the system at the collective coordinate ${\bm q}$. The corresponding collective wave function $g_{\bm q}(t)$ is defined by the transformation \cite{Reinhard1987RPP}
\begin{equation}\label{Eq_f_g}
g=\mathcal{N}^{1/2}f,
\end{equation}
where $\mathcal{N}^{1/2}$ is the square root of the overlap kernel matrix. Inserting Eq.~(\ref{Eq_f_g}) into Eq.~(\ref{TD-HW-f}), the time evolution of the collective wave function is governed by the equation \cite{Regnier2019PRC}
\begin{equation}
\label{Eq_HW_4}
    i\hbar \dot{g}=\mathcal{N}^{-1/2}(H-H^{MF})\mathcal{N}^{-1/2}g+i\hbar\dot{\mathcal{N}}^{1/2}\mathcal{N}^{-1/2}g.
\end{equation}
%-------------------------------------------------------------------------------
\subsection{Observables $\hat{O}$}
%-------------------------------------------------------------------------------
The kernel of any observable $\hat{O}$
\begin{equation}
 \mathcal{O}_{\bm {q'q}}=\langle\Phi_{q'}(t)|\hat{O}|\Phi_q(t)\rangle
\end{equation}
can be mapped to the corresponding collective operator $\mathcal{O}^c$:
\begin{equation}
\mathcal{O}^c=\mathcal{N}^{-1/2}\mathcal{O}\mathcal{N}^{-1/2}.
\end{equation}
The expectation value of the observable $\hat{O}$ in the correlated GHW state reads
\begin{equation}
\label{Eq_observable}
\langle \Psi(t)|\hat{O}|\Psi(t)\rangle =f^\dag \mathcal{O} f=g^\dag \mathcal{O}^c g.
\end{equation}

%-------------------------------------------------------------------------------
\section{Numerical details}\label{sec_numerics}
%-------------------------------------------------------------------------------
In the present calculation that employs the generalized TD-GCM, the mesh spacing of the lattice is 1.0 fm for all directions, and the box size is $L_x\times L_y\times L_z=20\times20\times60~{\rm fm}^3$.
The time-dependent Dirac equation \eqref{Eq_td_Dirac_eq_BCS} is solved using the predictor-corrector method, and the time-dependent equations \eqref{Eq_td_nkapp_eq_BCS} and \eqref{Eq_HW_4} with the Euler algorithm.
The step for the time evolution is $6.67\times10^{-4}$~zs. The point-coupling relativistic energy density functional PC-PK1 \cite{Zhao2010PRC} is used in the particle-hole channel, together with a monopole pairing interaction.
The pairing strength parameters: $-0.135$ MeV for neutrons, and $-0.230$ MeV for protons, are determined by the empirical pairing gaps of $^{240}$Pu,
using the three-point odd-even mass formula~\cite{Bender2000pairing}.
The initial states for the TD-DFT are obtained by self-consistent deformation-constrained relativistic DFT calculations in a three-dimensional lattice space, using the inverse Hamiltonian and Fourier spectral methods~\cite{ren17dirac3d, ren19LCS, ren20_NPA}, with the box size: $L_x\times L_y\times L_z=20\times20\times50~{\rm fm}^3$.

%-------------------------------------------------------------------------------
\section{Induced fission of $^{\bf{240}}$Pu}
\label{sec_240Pu}
%-------------------------------------------------------------------------------

In the analysis of induced fission  in Ref.~\cite{Ren2022a}, we used a consistent microscopic framework, based on TD-GCM and TD-DFT, to model the entire process of induced fission. The TDGCM was used to evolve adiabatically a set of quadrupole and octupole deformation degrees of freedom of $^{240}$Pu from the quasi-stationary initial state to the outer barrier and beyond. Starting from an iso-energy contour beyond the outer barrier, for which the probabilities that the collective wave function reaches these points at any given time were calculated using TDGCM, the TDDFT was subsequently used to model the dissipative fission dynamics in the saddle-to-scission phase. It has been shown, however, that the time-dependent GCM+DFT approach cannot reproduce both the distributions of fission yields and kinetic energy at a quantitative level. The principal reason is that quantum fluctuations, that are essential for a quantitative estimate of fission yields, are included in GCM but not in DFT trajectories that model the saddle-to-scission evolution of the fissioning system.

\begin{figure}[]
\centering
\includegraphics[width=1.0\textwidth]{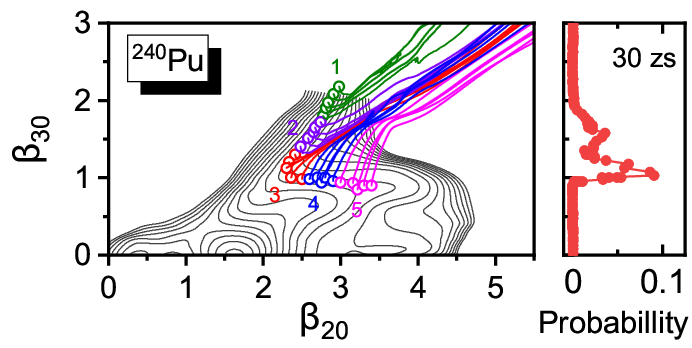}
\caption{(color online). Self-consistent deformation energy surface of $^{240}$Pu in the plane of quadrupole-octupole axially-symmetric deformation parameters, calculated with the relativistic density functional PC-PK1 and a monopole pairing interaction. Contours join points on the surface with the same energy, and the open dots correspond to points on the iso-energy curve at 1 MeV below the energy of the equilibrium minimum. In the panel on the right, the normalized probability that the initial TD-GCM wave packet reaches a particular point after 30 zs, is plotted as function of the octupole deformation parameter. The curves correspond to self-consistent TD-DFT fission trajectories from the initial points, to be used as generator states of the generalized TD-GCM.}
 \label{fig:240Pu}
\end{figure}

In the present study, we aim to improve the description of the saddle-to-scission phase of the fission process, by performing generalized time-dependent GCM calculations that use a basis of, generally non-orthogonal and overcomplete, TD-DFT fission trajectories to build the correlated TD-GCM wave function. As in the first part of this work \cite{Li2023PRC} and in Ref.~\cite{Ren2022a}, we consider the illustrative case of fission of $^{240}$Pu. In Fig.~\ref{fig:240Pu}, we plot the two-dimensional microscopic self-consistent mean-field deformation energy surface of $^{240}$Pu as a function of the axial quadrupole ($\beta_{20}$) and octupole ($\beta_{30}$) deformation parameters. The open dots on the energy surface correspond to the iso-energy contour at 1 MeV below the energy of the equilibrium minimum, located beyond the outer fission barrier.
The panel on the right displays the normalized probability distribution that the initial TD-GCM wave packet reaches a  particular point after a time of $30$ zs, as a function of $\beta_{30}$. These points are used as initial locations for the TD-DFT calculations that produce a set of self-consistent fission trajectories to be used as generator states of the generalized TD-GCM. We have plotted 25 trajectories that start from points on the iso-energy contour, reached by the initial collective wave function with non-negligible probability.

In principle, one proceeds by building the correlated TD-GCM wave function in the basis of all DFT fission trajectories. At present, however, in most cases this is computationally prohibitive. Therefore, in this illustrative calculation, we have grouped the DFT fission trajectories in sets of 5 neighbouring trajectories, denoted by different colours in Fig.~\ref{fig:240Pu}. Of course, this is not an ideal solution as trajectories from different sets might cross, but will suffice to illustrate the effect of including quantum fluctuations in the description of self-consistent evolution of the fissioning system.

\begin{figure}[]
\centering
\includegraphics[width=0.5\textwidth]{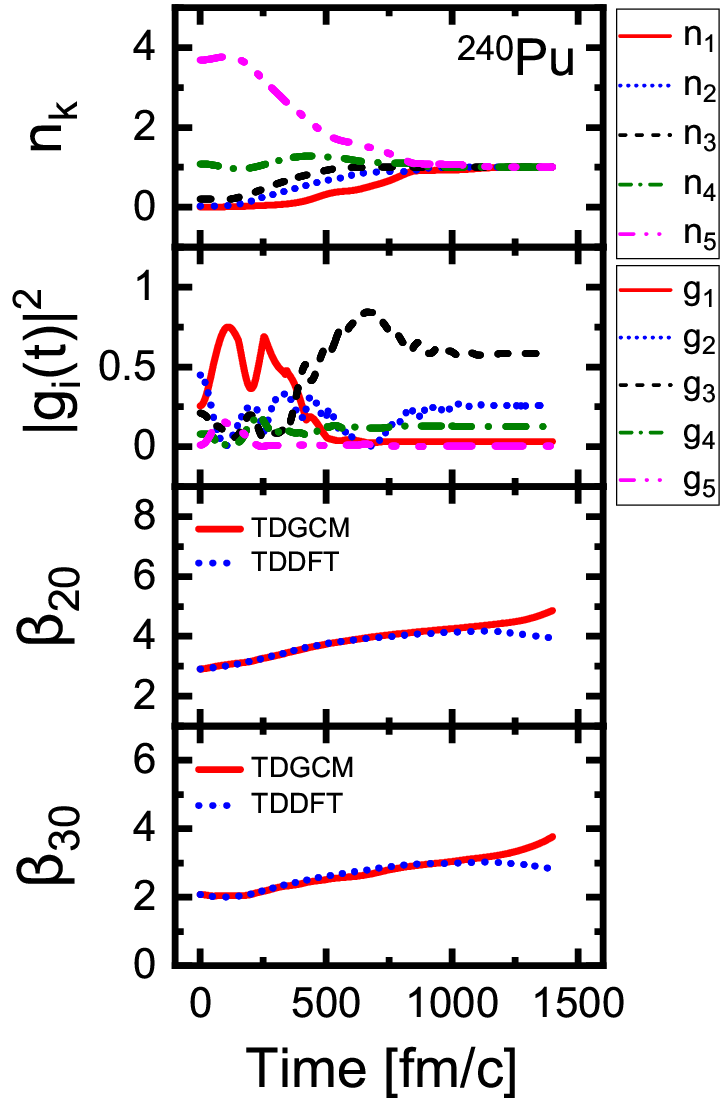}
\caption{(color online). Fission of $^{240}$Pu with the five TD-DFT trajectories that start in region $1$. The initial point for the generalized GCM evolution is at $\beta_{20}=2.91$ and $\beta_{30}=2.08$ on the deformation energy surface. The top panel displays the eigenvalues of the overlap kernel, and the square moduli of components of the TD-GCM collective wave function are shown in the second panel.
The time evolution of the quadrupole and octupole deformations on the way to scission and beyond, is compared to the single TD-DFT trajectory starting from the same point, in the two lower panels.
}
 \label{fig:Pu240-tdgcm-region1-2}
\end{figure}

We start by considering the time-evolution of the correlated GCM wave function that starts at the point $\beta_{20}=2.91$ and $\beta_{30}=2.08$ on the deformation energy surface, and is a superposition of 5 TD-DFT trajectories from region 1 (green trajectories in Fig.~\ref{fig:240Pu}). When these trajectories are used as generator states of the generalized TD-GCM, their initial overlaps are relatively large, as shown by the eigenvalues of the overlap kernel in the top panel of Fig.~\ref{fig:Pu240-tdgcm-region1-2}. One notices that the eigenvalues approach 1 asymptotically with time, which means that the trajectories become orthogonal. This is because after scission they correspond to distinct pairs of fragments with different particle numbers. The second panel of Fig.~\ref{fig:Pu240-tdgcm-region1-2} displays the evolution of the five components of the collective TD-GCM wave function. One notices considerable mixing and, therefore, pronounced quantum fluctuations in the period before scission, while the contributions of the square moduli of components $g$ are constant after scission, reflecting the orthogonality of the TD-DFT basis trajectories.
In the lower two panels, we plot the time dependence of the quadrupole and octupole deformations on the way to scission and beyond, in comparison to the single TD-DFT trajectory that also starts at $\beta_{20}=2.91$ and $\beta_{30}=2.08$. The interesting result here is that, while the single TD-DFT trajectory eventually gets trapped in local minima and does not lead to scission ($\beta_{20}$ and $\beta_{30}$ decrease with time), quantum fluctuations induced by admixing configurations that correspond to the other four TD-DFT trajectories are sufficient to produce a scission event.

\begin{figure}[]
\centering
\includegraphics[width=0.6\textwidth]{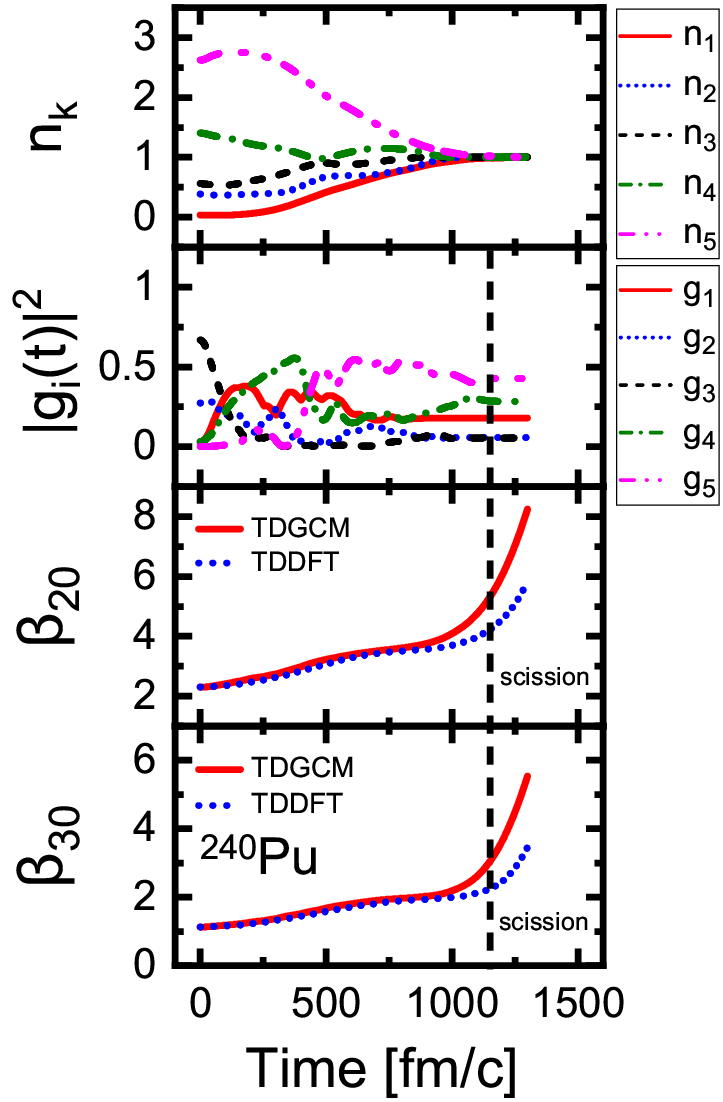}
\caption{(color online). Same as in the caption to Fig.~\ref{fig:Pu240-tdgcm-region1-2} but for the five TD-DFT trajectories that start in region $3$ on Fig.~\ref{fig:240Pu}. The initial point for the generalized GCM evolution is at $\beta_{20}=2.30$ and $\beta_{30}=1.13$ on the deformation energy surface. The vertical dashed line denotes the instant of scission.
}
 \label{fig:Pu240-tdgcm-region3-3}
\end{figure}

In the next example, we examine the superposition of $5$ fission trajectories that start in region $3$ on Fig.~\ref{fig:240Pu}.  This is, of course, not an optimal choice as obviously one expects pronounced admixtures of TD-DFT trajectories from regions 2 and 4 as well, but at present we cannot evolve in time a collective wave function expanded in a much larger TD-DFT basis. The evolution of the correlated GCM wave function starts at the point $(\beta_{20},\beta_{30})=(2.30,1.13)$, and the time dependence of the eigenvalues of the overlap kernel, the square moduli of components of the wave function, and the quadrupole and octupole deformations, are shown in Fig.~\ref{fig:Pu240-tdgcm-region3-3}. In this particular case, starting from $(\beta_{20},\beta_{30})=(2.30,1.13)$, both the TD-DFT and generalized TD-GCM trajectories lead to scission, though the increase of deformation parameters is obviously steeper for the latter. Scission occurs after approximately $1200$ fm/c, after which the TD-DFT basis trajectories are orthogonal and the eigenvalues of the overlap matrix are all equal to 1. At scission, each TD-DFT basis trajectory corresponds to a specific number of protons and neutrons in the fission fragments. This means that, at scission and beyond, the square moduli of components of the TD-GCM collective wave function, shown in the second panel of Fig.~\ref{fig:Pu240-tdgcm-region3-3}, denote the probabilities to observe the light and heavy fragments of the corresponding TD-DFT trajectory.

\begin{figure}[]
\centering
\includegraphics[width=0.6\textwidth]{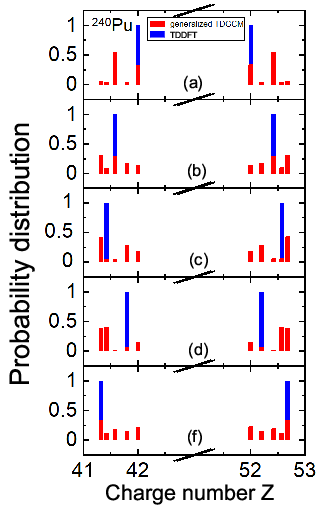}
\caption{(color online). Probability distributions of charge yields for the TD-DFT trajectories that start in region $3$ on the deformation energy surface in Fig.~\ref{fig:240Pu}. In each panel, from top to bottom, the blue bars denote the charge of the light and heavy fragments calculated for the TD-DFT trajectory that starts from the initial points: $(\beta_{20},\beta_{30})=(2.41,1.30), (2.33,1.20), (2.30,1.13), (2.36,1.00)$, and $(2.50,0.96)$, respectively. The red bars, normalized to 1 for the light and heavy fragments, correspond to the charge yields obtained for the generalized TD-GCM trajectory that starts from the same initial point as the TD-DFT one, but is a superposition of all five basis trajectories.
}
 \label{fig:charge_yield_3}
\end{figure}

This is illustrated in Fig.~\ref{fig:charge_yield_3}, where we display the probability distributions of charge yields for the TD-DFT trajectories that start in region $3$ on the deformation energy surface in Fig.~\ref{fig:240Pu}. In each panel, from top to bottom, the blue bars denote the charge of the light and heavy fragments calculated for the TD-DFT trajectory that starts from the initial point:
$(\beta_{20},\beta_{30})=(2.41,1.30), (2.33,1.20), (2.30,1.13), (2.36,1.00)$, and $(2.50,0.96)$, respectively. Obviously, for each TD-DFT trajectory one obtain just one charge number for the light, and one for the heavy fragment. Since these are average yields, that is, no number projection is performed, they are not necessarily integer numbers. In addition, the values are located in a small interval because all five TD-DFT trajectories start from neighbouring points on the deformation energy surface. The red bars, normalized to 1 for the light and heavy fragments, correspond to the charge yields obtained for the generalized TD-GCM trajectory that starts from the same initial point as the TD-DFT trajectory, but is a superposition of all five basis trajectories. In this case, a distribution of charge yields is obtained, rather than a single light and heavy fragment. This is a direct manifestation of quantum fluctuations that are intrinsic to the generalized TD-GCM.

\begin{figure}[]
\centering
\includegraphics[width=0.6\textwidth]{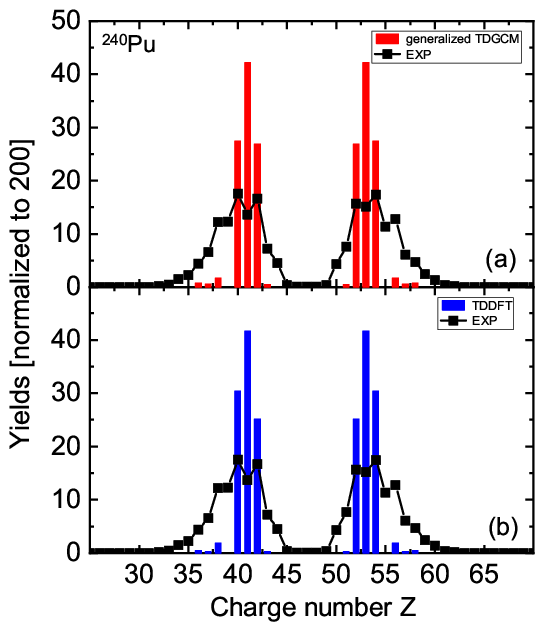}
\caption{(color online). Charge yields for induced fission of $^{240}$Pu. The yields calculated with the generalized TD-GCM (a), and TD-DFT (b) methods, are shown in comparison with the experimental charge distribution.
The data are from Ref.~\cite{Ramos2018PRC}, and correspond to an average excitation energy of 10.7 MeV.
}
 \label{fig:charge_yield_all}
\end{figure}

To calculate the total charge yields, we multiply each scission event by the probability that the initial GCM collective wave packet has reached the corresponding starting point of the generalized TD-GCM trajectory after $30$ zs (panel on the right of Fig.~\ref{fig:240Pu}). When the calculated fractional charge numbers are assigned to the nearest integer charge, and the yields are normalized to $\sum_{Z} Y(Z) = 200$, the resulting total charge yields are plotted in the upper panel of Fig.~\ref{fig:charge_yield_all}, in comparison to the experimental fragment charge distribution \cite{Ramos2018PRC}. The latter correspond to an average excitation energy of 10.7 MeV. Compared to the data, it is evident that the calculated yields reproduce the position of the peaks but not the tails of the experimental distribution. The reason are the very small probabilities that the initial GCM collective wave packet reaches the points on the iso-energy contour that would lead to the corresponding scission events in the tails of the charge distribution. This could change, however, if the GCM probabilities at the iso-energy curve were calculated at a finite temperature that correspond to the experimental excitation energy of 10.7 MeV. As shown in our studies of Refs.~\cite{Zhao2019PRC_1,Zhao2019PRC_2}, by employing the TD-GCM  framework based on a finite-temperature extension of nuclear density functional theory, theoretical yields are predicted in considerably better agreement with available data. In this work, however, we have included pairing correlations but not finite temperature effects.

Since each correlated TD-GCM trajectory is expanded in a basis of only five neighbouring TD-DFT trajectories, the total charge yields are not very different from those obtained using the standard TD-DFT method \cite{Ren2022a}, shown in the lower panel of Fig.~\ref{fig:charge_yield_all}. Taking into account the results displayed in Fig.~\ref{fig:charge_yield_3}, it is apparent that a much larger basis of TD-DFT trajectories is required for the generalized TD-GCM to produce quantitatively different charge yields.

\begin{figure}[]
\centering
\includegraphics[width=0.6\textwidth]{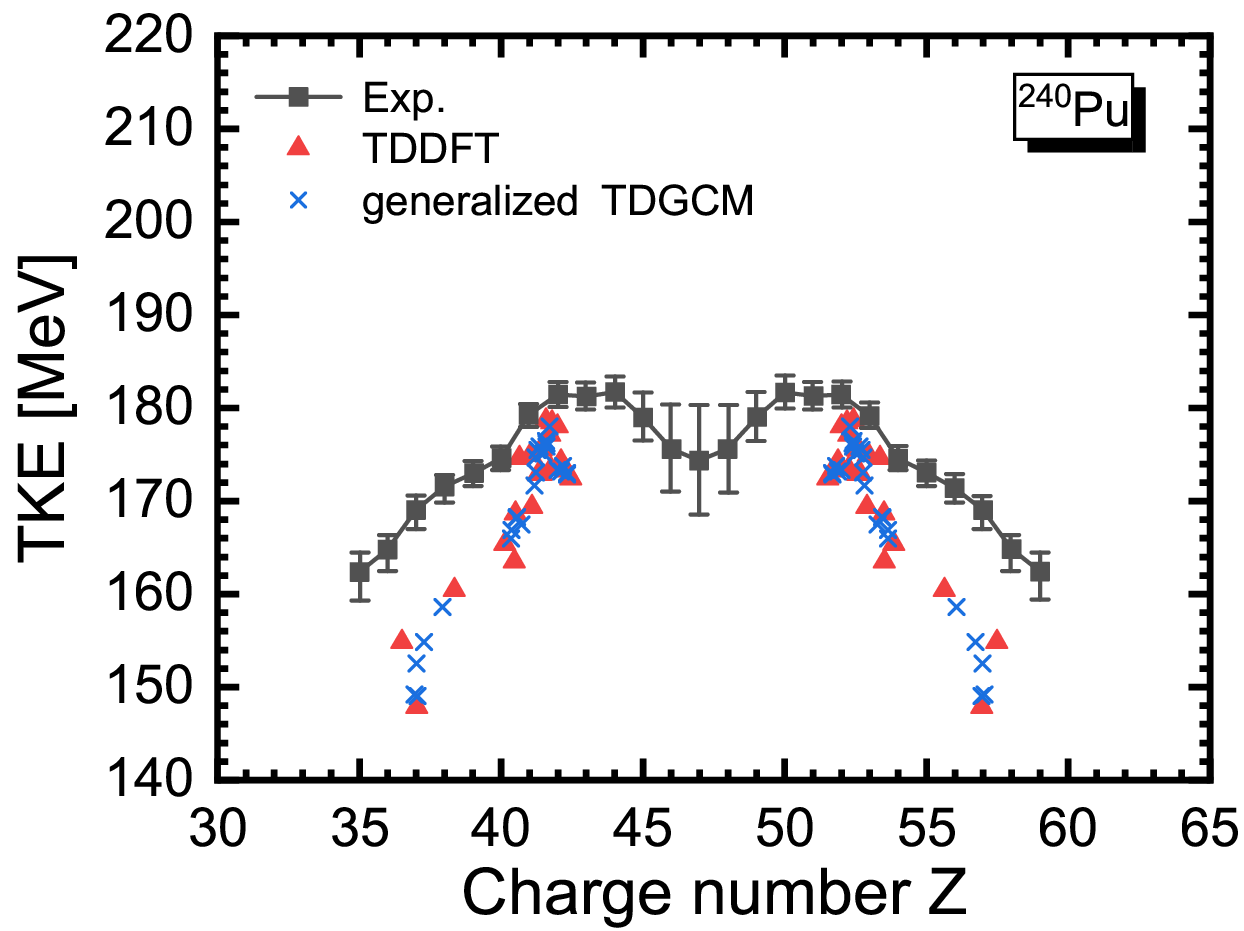}
\caption{(color online). The calculated total kinetic energies of the nascent fragments for induced fission of $^{240}$Pu, as functions of the fragment charge. The generalized TD-GCM and TD-DFT results are shown in comparison
to the data~\cite{Caamano2015PRC}.
}
 \label{fig:TKE}
\end{figure}

Similar results are also obtained when the TD-DFT and generalized TD-GCM are compared in a calculation of the total kinetic energy distribution, as shown in Fig.~\ref{fig:TKE}.
For one TD-DFT trajectory, the total kinetic energy (TKE) at a finite distance between the fission fragments ($\approx 25$ fm, at which shape relaxation brings the fragments to their equilibrium shapes) is calculated using the expression
\begin{equation}
E_{TKE}=\frac{1}{2}mA_{\rm H}^{\bm q}\bm{v}^2_{{\rm H},{\bm q}}+\frac{1}{2}mA_{\rm L}^{\bm q}\bm{v}^2_{{\rm L},{\bm q}}+E_{\rm Coul}^{\bm q},
\end{equation}
where the velocity of the fragment $f=H,L$ reads
\begin{equation}
\bm{v}_{f,{\bm q}}=\frac{1}{mA_{f}^{\bm q}}~\int_{V^{\bm q}_f} d\bm{r}~\bm{j}_{\bm q}(\bm{r}) \;,
\end{equation}
and $\bm{j}(\bm{r})$ is the total current density. The integration is over the half-volume corresponding to the fragment $f$, and $E_{\mathrm{Coul}}$ is the Coulomb energy. For the correlated TD-GCM state, the
charge number of the fragments and the total kinetic energy of a trajectory is calculated using Eq.~(\ref{Eq_observable}). In comparison with the experimental results \cite{Caamano2015PRC}, both methods qualitatively reproduce the experimental TKEs for the fragments close to the peaks of the charge yields distribution, but predict TKEs considerably below the experimental values for the tails of the distribution. As already noted in Ref.~\cite{Ren2022a}, this is due to the fact that the calculated TKEs do not include the contribution of pre-scission energy because the initial points for the fission trajectories are on the deformation energy surface, 1 MeV below the energy of the equilibrium minimum and, therefore, more than 10 MeV below the `physical' value. Also in this case, the inclusion of finite temperature effects could improve the agreement with experimental results.

\begin{figure}[]
\centering
\includegraphics[width=0.6\textwidth]{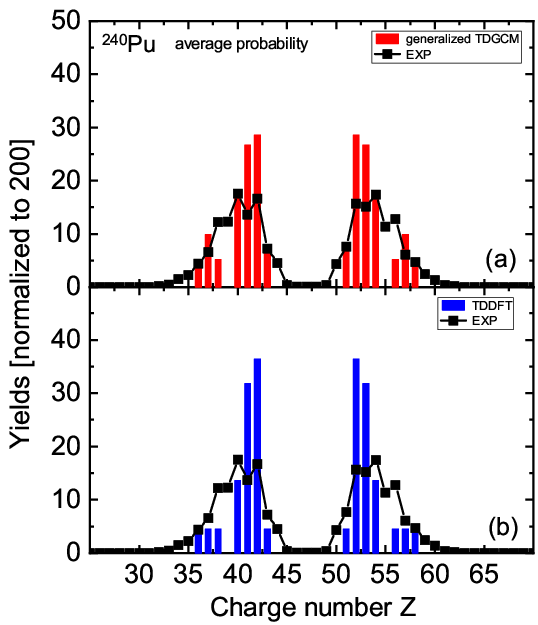}
\caption{(color online). Same as Fig.~\ref{fig:charge_yield_all}, but with equal initial probabilities for each fission  trajectory.
}
 \label{fig:charge_yield_average}
\end{figure}
To illustrate the possible effect of including finite temperature in the description of induced fission dynamics (cf. our recent study in Ref.~\cite{Li2023PRC_FT}), in Fig.~\ref{fig:charge_yield_average} we plot the generalized TD-GCM and TD-DFT charge yields, calculated in the same way as those shown in Fig.~\ref{fig:charge_yield_all}, but with equal probabilities for all 25 initial points. This is, of course, exaggerated as even at the temperature that corresponds to the experimental excitation energy, the initial points will not be reached with equal probability by the GCM collective wave packet after $30$ zs. However, we notice a rather pronounced effect of the initial probabilities and a much better agreement with experiment, both for the TD-DFT and generalized TD-GCM.

\begin{figure}[]
\centering
\includegraphics[width=0.6\textwidth]{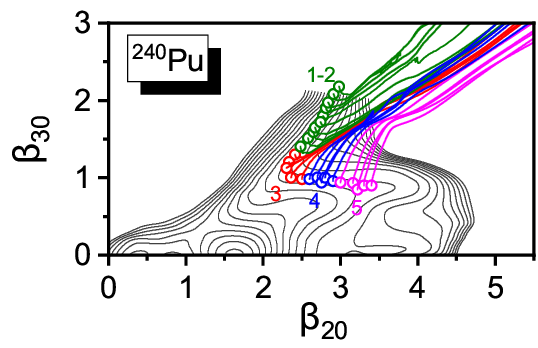}
\caption{(color online). Same as the left panel of Fig.\ref{fig:240Pu}, but the initial points in regions $1$ and $2$ are here merged into a larger basis $1$-$2$, which includes $10$ TD-DFT trajectories.
}
 \label{fig:240Pu_merge}
\end{figure}

As already noted in the discussion above, the choice to expand each generalized TD-GCM fission trajectory in a basis of only 5 TD-DFT trajectories is dictated by the available computational capabilities. While we hope that in the future it will become possible to perform more realistic calculations, here the effect is illustrated by merging the initial points in regions $1$ and $2$ into the basis $1-2$ with $10$ TD-DFT trajectories, as shown in Fig.~\ref{fig:240Pu_merge}.
The trajectories in region $1$ produce the fragments in the tail of the charge distribution (charge number $Z=36,37,38$, and $56,57,58$), while the trajectories in region $2$ mainly contribute to the peak fragments with charge number $Z=40,41$ and $53,54$.

\begin{figure}[]
\centering
\includegraphics[width=0.6\textwidth]{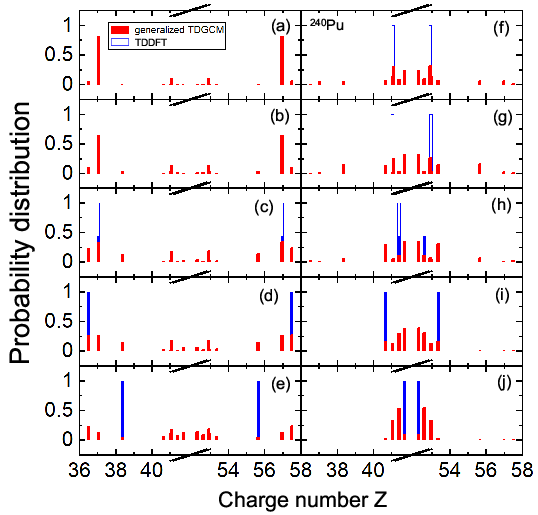}
\caption{(color online). Probability distributions of charge yields for the TD-DFT trajectories that start in region $1-2$ on the deformation energy surface in Fig.~\ref{fig:240Pu_merge}. In each panel, from (a) to (j), the blue bars denote the charge of the light and heavy fragments calculated for the TD-DFT trajectory that starts from an initial point in decreasing order of $\beta_{30}$, respectively. The red bars, normalized to 1 for the light and heavy fragments, correspond to the charge yields obtained for the generalized TD-GCM trajectory that starts from the same initial point as the TD-DFT trajectroy, but is a superposition of all ten basis trajectories.
}
 \label{fig:charge_yield_merge}
\end{figure}

Starting from one of the ten initial points in the region $1-2$, and expanding in the basis of 10 TD-DFT trajectories, the corresponding generalized TD-GCM charge yields are obtained. In each panel of Fig.\ref{fig:charge_yield_merge}, from (a) to (j), the blue bars denote the charge of the light and heavy fragments calculated for the TD-DFT trajectory that starts from an initial point in decreasing order of $\beta_{30}$, respectively. In panels (a) and (b) no blue bars are shown, and this means that the corresponding trajectory does not lead to scission and the formation of fission fragments. The red bars, normalized to 1 for the light and heavy fragments, correspond to the charge yields obtained for the generalized TD-GCM trajectory that starts from the same initial point as the TD-DFT trajectory, but is a superposition of all ten basis trajectories. We note that, even in the cases in which the TD-DFT trajectory does not end in scission (panels (a) and (b)), the correlated TD-GCM state starting from the same initial point, but expanded in a basis that includes neighbouring TD-DFT trajectories, produces a broad distribution of fission fragments. When compared with the yields shown in Fig.~\ref{fig:charge_yield_3}, where we have shown results obtained with the 5 TD-DFT basis trajectories in region $3$, the advantage of expanding the generalized TD-GCM state in a larger basis is obvious, as it results in a much broader distribution.

%-------------------------------------------------------------------------------------------------------------
\section{Summary}\label{sec_summ}
%-------------------------------------------------------------------------------------------------------------
In this work, we have extended the generalized TD-GCM framework, developed in Ref.~\cite{Li2023PRC}, to include also pairing correlations. The correlated GCM wave function of the nuclear system is determined by time-dependent generator states and weight functions. In the particular implementation used in Ref.~\cite{Li2023PRC} and in this work, the initial generator states are calculated using the deformation-constrained self-consistent mean-field method, and are evolved in time by the standard equations of density functional theory (DFT). The DFT generator states form a time-dependent basis in which the GCM wave function is expanded.

The particle-hole channel of the interaction is determined by a Hamiltonian derived from a relativistic (covariant) energy density functional. Here we have employed the point-coupling functional PC-PK1 \cite{Zhao2010PRC}. Pairing correlations are included in a standard BCS approximation, with time-dependent pairing tensor and occupation probabilities, and a monopole pairing force is used, with strength parameters for protons and neutrons determined by empirical pairing gaps.  The inclusion of pairing allows for a more complete modelling of various time-dependent phenomena, but is technically much more complicated and, therefore, computationally demanding.

One of the main motivation for expanding the GCM framework with time-dependent generator states, and also for including pairing correlations, is the possibility for a more realistic description of the final, saddle-to-scission, phase of the process of induced fission. To take into account the dissipation of energy of nuclear collective motion into excitation of intrinsic degrees of freedom, that is, the heating of the system as it evolves toward scission, the dynamics of this phase of fission is usually modelled by methods based on time-dependent density-functional theory. However, since TD-DFT based models only describe a single fission trajectory by propagating nucleons independently from a given initial point on the deformation energy surface, they do not take into account quantum fluctuations that arise from a coherent superposition of different trajectories. In the generalized TD-GCM, with TD-DFT trajectories that start from different points on the deformation energy surface as time-dependent basis states in which the correlated nuclear function is expanded, the description of fission dynamics goes beyond the adiabatic approximation of the standard GCM, and also includes quantum fluctuations.

In the present illustrative study, we have applied the new model to the description of saddle-to-scission dynamics of induced fission of $^{240}$Pu. The charge yields and total kinetic energy distribution, obtained by the generalized TD-GCM,  have been compared to those calculated using the standard TD-DFT approach, and with available data. Although very encouraging results have been obtained for the principal charge yields and the corresponding kinetic energies, a detailed comparison with data indicates the necessity to expand the GCM correlated function in a relatively large basis of TD-DFT fission trajectories. This is, at present, still computationally prohibitive and will require the utilisation of large computing facilities and/or more efficient algorithms. In addition, to take into account the initial excitation energy of the fissioning system, the theoretical formalism will have to be extended to finite temperatures. Work along these lines is in progress and will be reported in a forthcoming publication.

\begin{acknowledgments}
This work has been supported in part by the High-end Foreign Experts Plan of China,
National Key R\&D Program of China (Contracts No. 2018YFA0404400),
the National Natural Science Foundation of China (Grants No. 12070131001, 11875075, 11935003, 11975031, and 12141501),
the High-performance Computing Platform of Peking University,
the QuantiXLie Centre of Excellence, a project co-financed by the Croatian Government and European Union through the European Regional Development Fund - the Competitiveness and Cohesion Operational Programme (KK.01.1.1.01.0004),
 and the Croatian Science Foundation under the project Uncertainty quantification within the nuclear energy density framework (IP-2018-01-5987).
\end{acknowledgments}

\clearpage
\bigskip
\bibliography{Pu240_fission}
%\begin{thebibliography}{99}

%\bibitem{Regnier2019PRC}

%\end{thebibliography}


\end{document}